\documentclass[twoside]{dis07}
\usepackage[latin1]{inputenc}
\usepackage[dvips]{graphicx,epsfig,color}
\usepackage{wrapfig,rotating}
\usepackage{amssymb,amsmath,array}

\newcommand{\sqrts}{\sqrt{s}}

\def\barr#1{\overline{\mbox{#1}}}

\def\ttt#1{\texttt{\small #1}}

\newcommand{\jpsi}{J/\psi}

\newcommand{\ups}{\Upsilon}

\newcommand{\pom}{I\!\!P}
\newcommand{\gaga}{\gamma\,\gamma}
\newcommand{\gp}{\gamma\mbox{-p}}
\newcommand{\gA}{\gamma\mbox{-A}}

\providecommand{\elel}{e^+e^-}
\providecommand{\mumu}{\mu^+\mu^-}
\providecommand{\lele}{l^{+}\,l^{-}}

\begin{document}
\title{Forward Physics at the LHC}


\author{David d'Enterria
\vspace{.3cm}\\
CERN, PH-EP, CH-1211 Geneva 23, Switzerland
}


\maketitle

\begin{abstract}
Small-angle detectors at the LHC give access to a broad physics programme 
within and beyond the Standard Model (SM). We review the capabilities of 
ALICE, ATLAS, CMS, LHCb, LHCf and TOTEM for forward physics studies in 
various sectors: soft and hard diffractive processes, exclusive Higgs production, 
low-$x$ QCD, ultra-high-energy cosmic-rays, and electro-weak measurements~\cite{url}.
\end{abstract}

\section*{Introduction}

\begin{figure}[htbp]
\centerline{
\includegraphics[width=0.45\columnwidth]{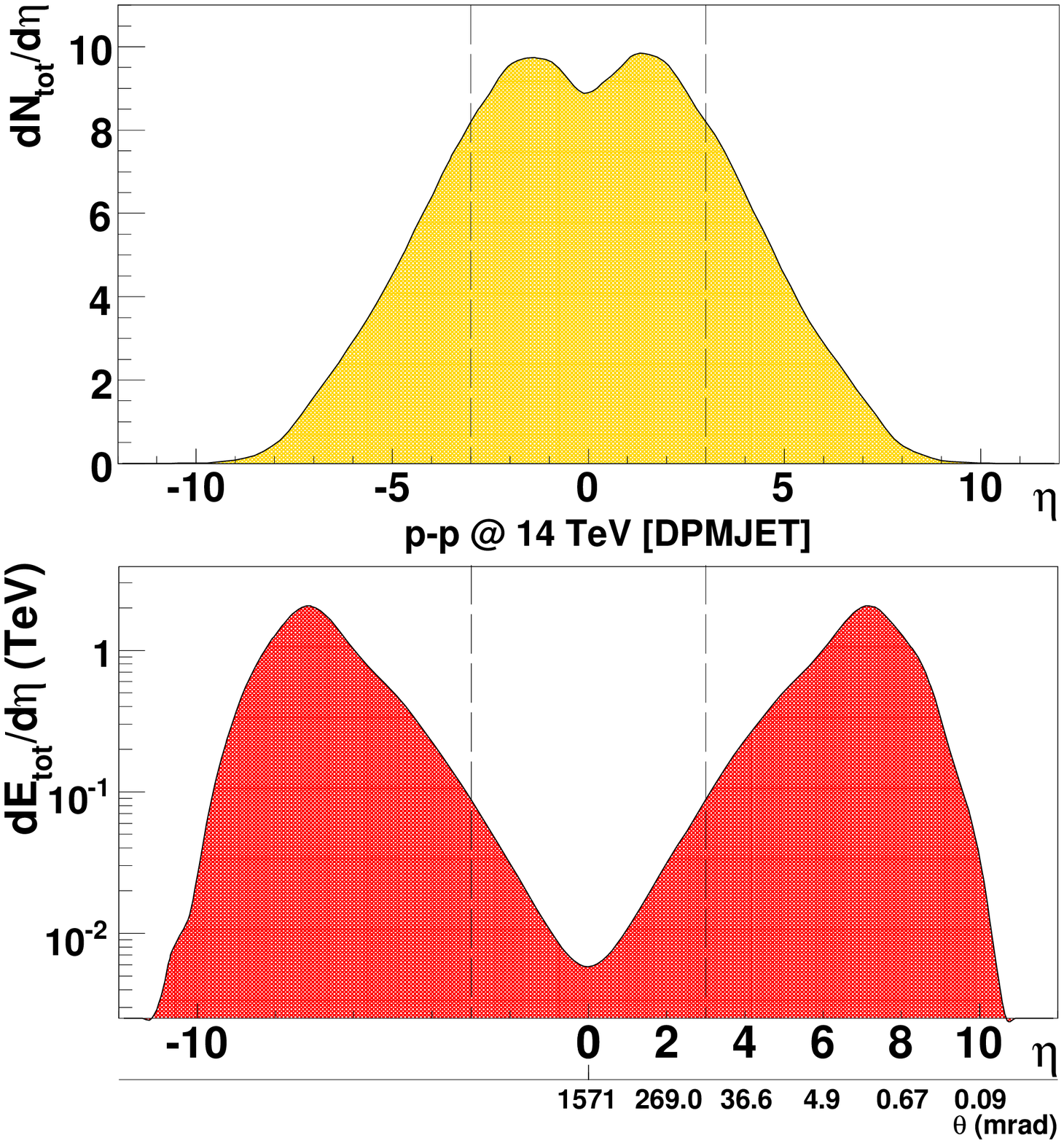}
\includegraphics[width=0.55\columnwidth]{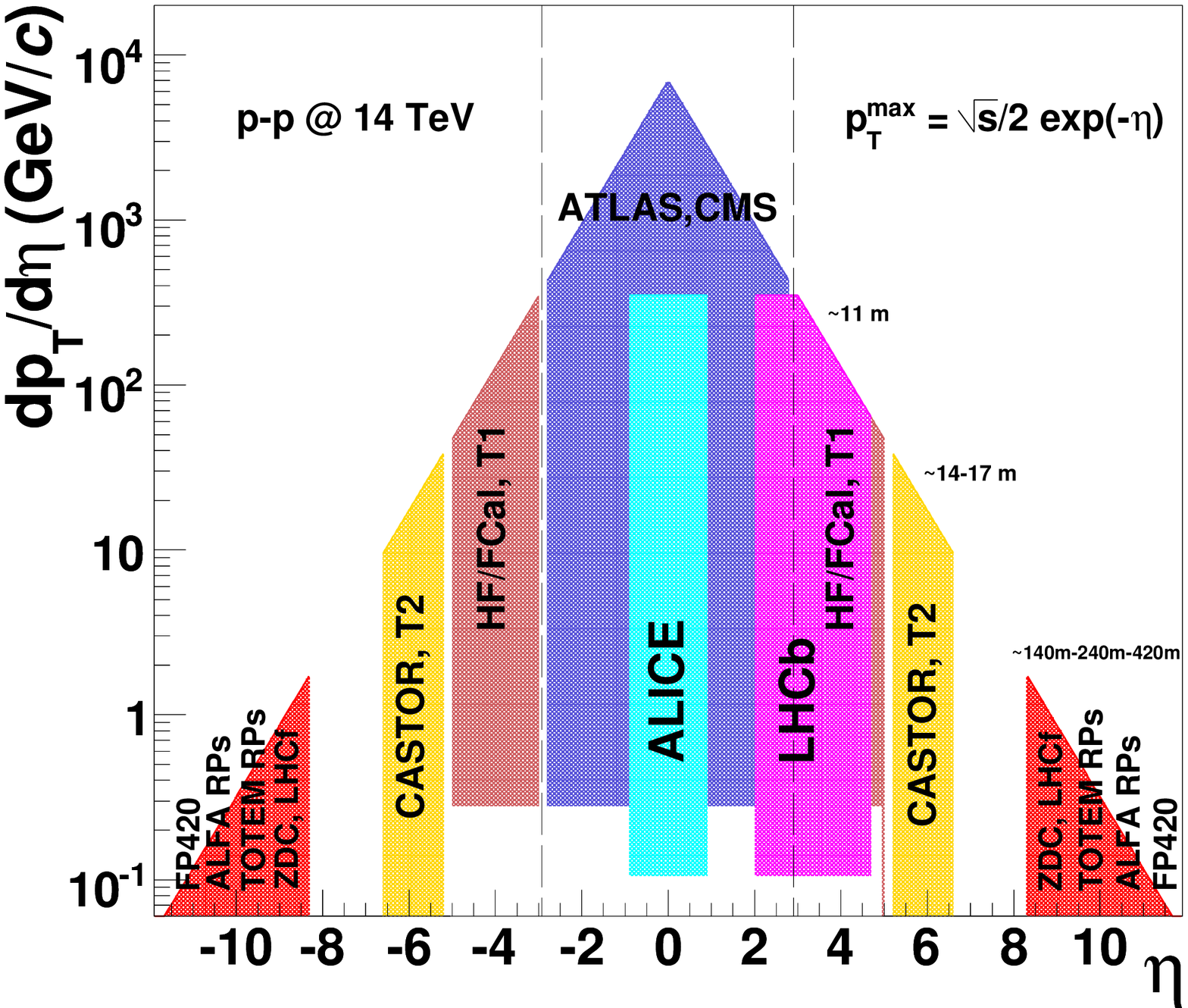}
}
\caption{Left: Pseudo-rapidity distributions for the total hadron multiplicity (top) and 
energy (bottom) in p-p at 14 TeV as given by the DPMJET3 model~\cite{dpmjet3}.
Right: Approximate $p_T$-$\eta$ coverage of current (and proposed) detectors at the LHC 
(adapted from~\cite{risto06}).}
\label{fig:dN_dE_deta}
\end{figure}

The CERN Large Hadron Collider (LHC) will provide the highest energy proton-proton 
and ion-ion collisions in the lab to date. The multi-TeV energy of the colliding beams
opens up a phase space for particle production in an unprecedented range spanning 
$\Delta\eta\sim$ 20 units of rapidity\footnote{The rapidity can be thought of as the 
relativistically-invariant measure of longitudinal velocity. Often the {\it pseudorapidity} 
$\eta$ = -ln$\,$tan($\theta/2$) which depends only on the polar angle 
wrt the beam axis, is used instead.}: $y_{\rm beam}$ = acosh($\sqrts/2$)= 9.54 for 
p-p at 14 TeV. As a general feature, particle production in hadronic collisions is peaked at central 
rapidities ($|y|\lesssim$ 3 at the LHC), whereas most of the energy is emitted at very low angles 
(Fig.~\ref{fig:dN_dE_deta} left).

\begin{figure}[htbp]
\centerline{
\includegraphics[width=0.95\columnwidth,height=6.cm]{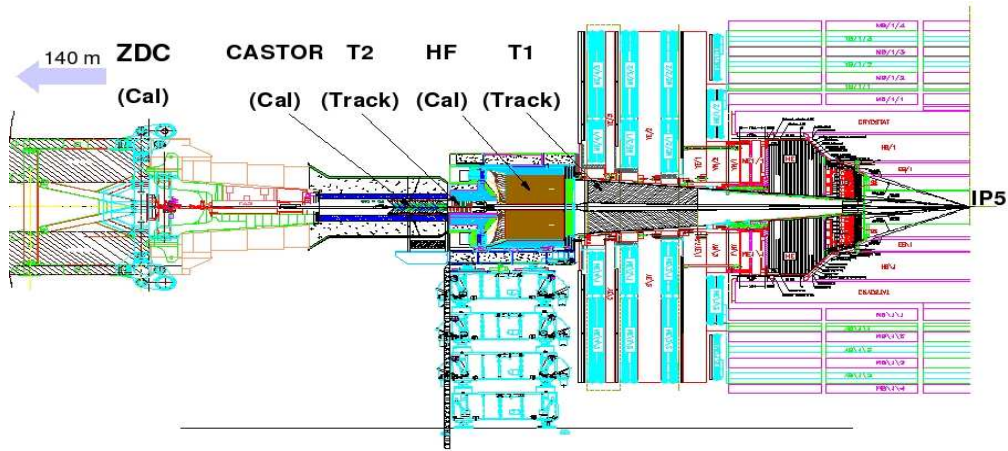}
}
\caption{Layout of the detectors in the CMS/TOTEM forward region~\cite{Albrow:2006xt}.}
\label{fig:fwd_cms}
\end{figure}

\noindent
All LHC experiments feature detection capabilities at forward rapidities without parallel
compared to previous colliders (Fig.~\ref{fig:dN_dE_deta} right, and Fig.~\ref{fig:fwd_cms}):
\begin{itemize}
\item ATLAS~\cite{atlas,Ask:2007fr} and CMS~\cite{Albrow:2006xt,cms,borras} not 
only cover the largest $p_T$-$\eta$ ranges at mid-rapidity for hadrons, electrons, photons
and muons, but they feature extended instrumentation at distances far away from 
the interaction point (IP): $\pm$11 m (ATLAS FCal and CMS HF hadronic calorimeters), 
$\pm$14~m (CMS CASTOR sampling calorimeter~\cite{castor}), 
$\pm$140~m (Zero-Degree-Calorimeters, ZDCs~\cite{Ask:2007fr,Grachov:2006ke}), and
$\pm$240~m (ATLAS Roman Pots, RPs~\cite{Ask:2007fr}).
\item ALICE~\cite{alice} and LHCb~\cite{lhcb} have both forward muon spectrometers 
in regions, $2\lesssim~\eta~\lesssim~5$, not covered by ATLAS or CMS. (In addition, 
ALICE has also ZDCs at $\pm$116 m~\cite{alice_zdc}).
\item The TOTEM experiment~\cite{totem}, sharing IP5 with CMS, features two types 
of trackers (T1 and T2 telescopes) covering $3.1<|\eta|<4.7$ and $5.2<|\eta|<4.7$ 
respectively, plus proton-taggers (Roman Pots) at $\pm$147 and $\pm$220 m.
\item The LHCf~\cite{lhcf} tungsten-scintillator/silicon calorimeters share the location 
with the ATLAS ZDCs $\pm$140 m away from IP1.
\item The FP420 R\&D collaboration~\cite{fp420,fp420b} aims at installing proton taggers at 
$\pm$420 m from both ATLAS and CMS IPs.
\end{itemize}

\noindent
Near beam instrumentation provides access to a rich variety of physics measurements
when used in three possible modes: (i) as detectors 
to {\it directly measure} a given final-state produced in the reaction 
(e.g. a jet in CASTOR, or a photon in LHCf/ZDC), (ii) as {\it tagging} 
devices for the (diffractively or elastically) scattered protons (in Roman Pots or other p-taggers),
and/or (iii) as {\it vetoing} devices of final-state particles produced in the collision (e.g. requiring 
no hadronic activity within a given rapidity range covered by one or more detectors).\\

\noindent
The following forward physics topics will be discussed in this short review:
\begin{enumerate}
\item {\bf Diffraction} (soft and hard) and {\bf elastic} scattering~\cite{Arneodo:2005kd,Goulianos:2007xr}. 
Measurements like the total p-p cross section, 
the rapidity-gap survival probability, and hard diffraction cross sections (heavy-Q, dijets, vector-bosons, ...) 
are accessible with the TOTEM and ALFA Roman Pots and/or by requiring a large enough 
rapidity gap in one (or both) of the forward hemispheres (e.g. HF+CASTOR in CMS).
\item  {\bf Central exclusive} production of the {\bf Higgs} boson and other heavy (new) 
particles~\cite{DeRoeck:2002hk,Khoze:2007td} can be studied combining the FP420 proton-taggers 
with the central ATLAS and CMS detectors.
\item The phenomenology of {\bf low-$x$ QCD} -- parton saturation, non-linear QCD evolution, 
small-$x$ PDFs, multi-parton scattering~\cite{raju,d'Enterria:2006nb} -- can be studied via
the measurement of hard QCD cross sections in the forward direction (e.g. jets, 
direct-$\gamma$ in HF/FCal, CASTOR, ...) or in exclusive photoproduction processes 
($\gp$, $\gA$ interactions)~\cite{Baltz:2007hw} tagged with forward protons (neutrons) in RPs (ZDCs).
\item Models of hadronic interactions of {\bf ultra-high-energy} (UHE) {\bf cosmic-rays} 
in the upper atmosphere~\cite{Engel:2005gf} can be effectively tuned by 
measuring in CASTOR, TOTEM, LHCf and ZDCs, the energy ($dE/d\eta$) 
and particle ($dN/d\eta$) flows in p-p, p-A, and A-A collisions.
\item {\bf Electroweak} interactions: Ultrarelativistic protons and ions 
generate fluxes of (equivalent) photons which can be used for a rich
programme of photoproduction studies at TeV energies~\cite{Baltz:2007hw}. 
Photon-induced interactions, tagged with forward protons (neutrons) in the RPs (ZDCs), 
allow one e.g. to measure the beam luminosity 
(via the pure QED process $\gaga \rightarrow \lele$) or to study (anomalous) gauge 
boson couplings (via $\gp, \gA \rightarrow$ p n $W$, or $\gaga \rightarrow ZZ,WW$).
\end{enumerate}

\section{Total and elastic cross sections}

\noindent
The measurement at the LHC of the total p-p cross section and $\rho$-parameter 
(ratio of real to imaginary part of the forward elastic scattering amplitude) 
provides a valuable test of fundamental quantum mechanics relations~\cite{Bourrely:2005qh} 
like the Froissart bound $\sigma_{\rm tot}<${\small{\it Const}} $\ln^2s$, the optical theorem 
$\sigma_{\rm tot}\sim Im f_{el}(t=0)$, and dispersion relations $Re f_{el}(t=0)\sim Im f_{el}(t=0)$. 

\begin{figure}[htbp]
\centering
\includegraphics[width=0.49\columnwidth,height=4.8cm]{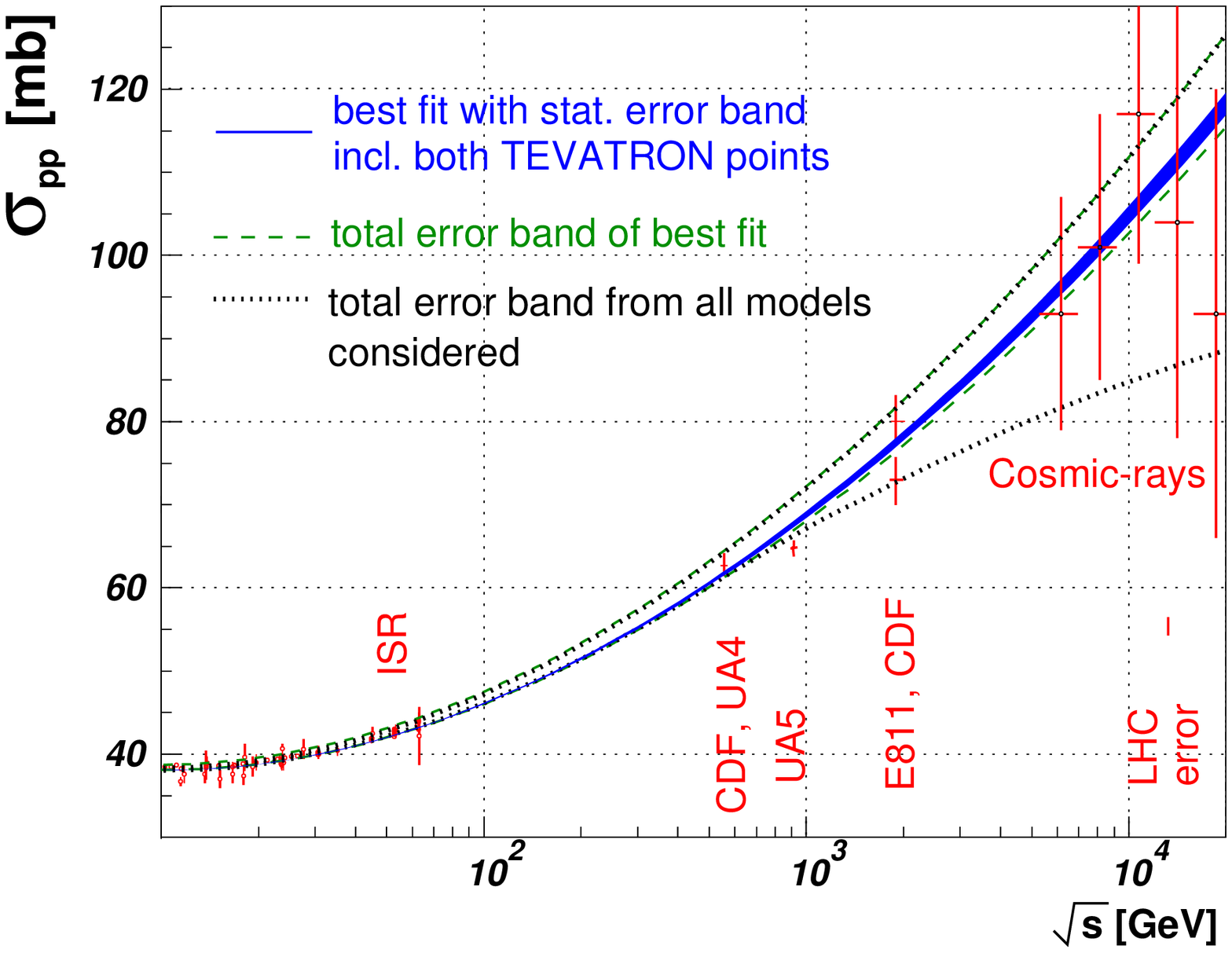}
\includegraphics[width=0.49\columnwidth,height=5cm]{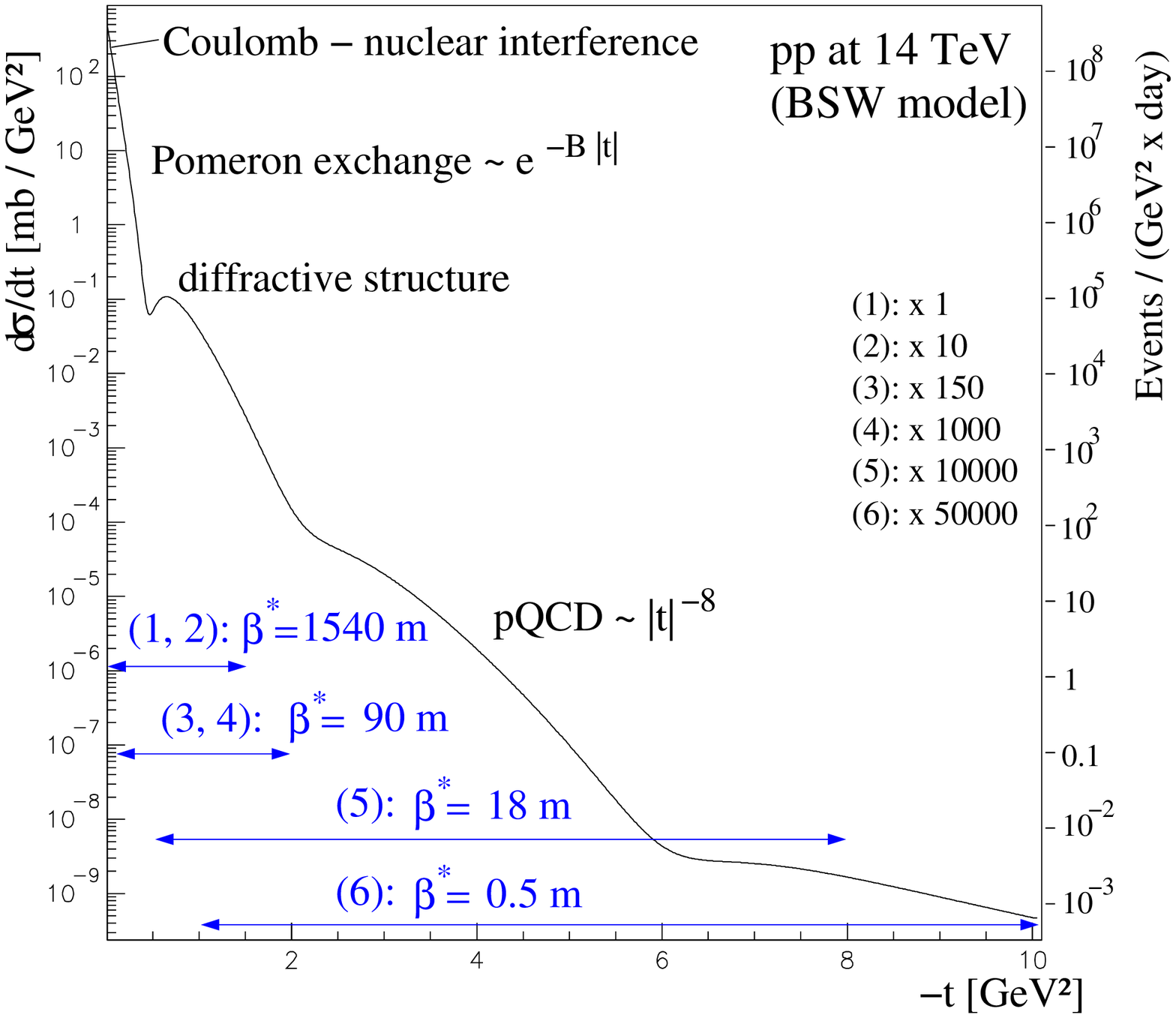}
\caption{Left: COMPETE predictions~\protect\cite{compete} for $\sigma_{tot}$ with statistical 
(blue solid) and total (dashed) errors (including the Tevatron ambiguity) compared to existing
data. Right: Prediction for elastic p-p scattering at the LHC with 
various beam optics settings~\protect\cite{Anelli:2006cz}.}
\label{fig:totem}
\end{figure}

\noindent
The main goal of TOTEM is to obtain a precise ($\sim 1$\%) measurement of the total and elastic 
p-p cross section over a large range of (low) 4-momentum transfers $-t\approx p^2\theta^2$
(Fig.~\ref{fig:totem}). The COMPETE~\cite{compete} extrapolation values of 
$\sigma_{\rm tot}$ = 111.5~mb and $\rho$ = 0.1361 at the LHC are uncertain to within 
$^{+5\%}_{-8\%}$ due to a 2.6$\sigma$ disagreement between the E710 and CDF 
measurements at Tevatron. In addition, TOTEM can also provide (via the optical theorem) 
the absolute p-p beam luminosity with reduced uncertainties using a low-$\beta$ setting.


\section{Diffractive physics}

\begin{figure}[htbp]
\centerline{
\includegraphics[width=0.99\columnwidth]{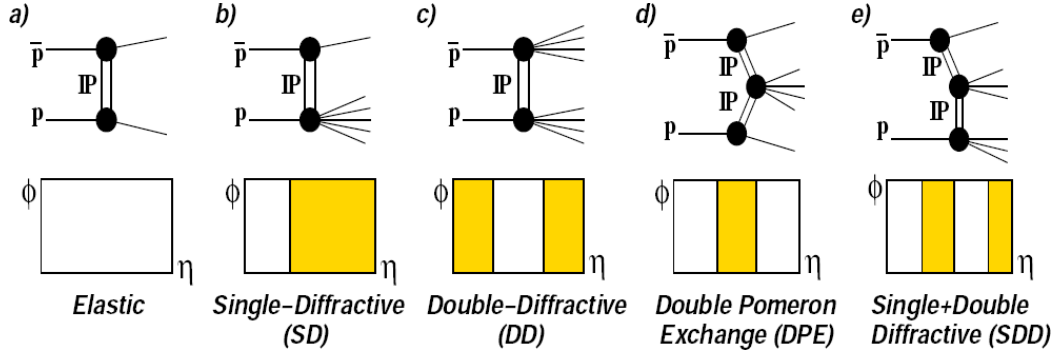}
}
\caption{Event topologies in $\eta$ vs azimuth $\phi$ for elastic and diffractive
p-p interactions. Shaded (empty) areas represent particle production (rapidity gaps) 
regions~\cite{Convery:2004jn}.}
\label{fig:diffract_diags}
\end{figure}

\noindent
Diffractive physics covers the class of interactions that contain large rapidity gaps 
(LRGs, $\Delta\eta\gtrsim$ 4) without hadronic production. Such event topologies imply
colorless exchange, requiring two or more gluons in a color-singlet state (a {\em Pomeron}, 
$\pom$). Depending on the number and relative separation of the LRGs, one further 
differentiates between single, double, or double-Pomeron-exchange (DPE) processes 
(Fig.~\ref{fig:diffract_diags}).

\noindent
On the one hand, soft diffraction processes are controlled by non-perturbative (Regge) dynamics 
and constitute a significant fraction ($\sim$20\%) of the total inelastic p-p cross section. 
Their characterization is thus important in order to determine the pile-up backgrounds at high 
luminosities. On the other, hard diffraction processes involve the production of a high-mass 
or large-$p_T$ state (X = Q$\barr{Q}$, jets, $W$, $Z$ ...) and are in principle calculable 
perturbatively by means of the factorisation theorem and diffractive (or generalised) 
Parton Distribution Functions, dPDFs (GPDs). The apparent breakdown of pQCD factorization 
in hard diffractive processes -- supported by a reduced gap-survival probability in Tevatron 
p-$\barr{p}$ compared to e-p at HERA -- has important phenomenological implications 
for LHC~\cite{Arneodo:2005kd,Goulianos:2007xr}. Of particular interest are hard 
exclusive DPE processes where the centrally produced system can be a new heavy 
particle 
(see Fig.~\ref{fig:DPE} and next Section).

\section{Higgs (and new) physics}

\begin{figure}[htbp]
\centerline{
\includegraphics[width=0.49\columnwidth,height=4.5cm]{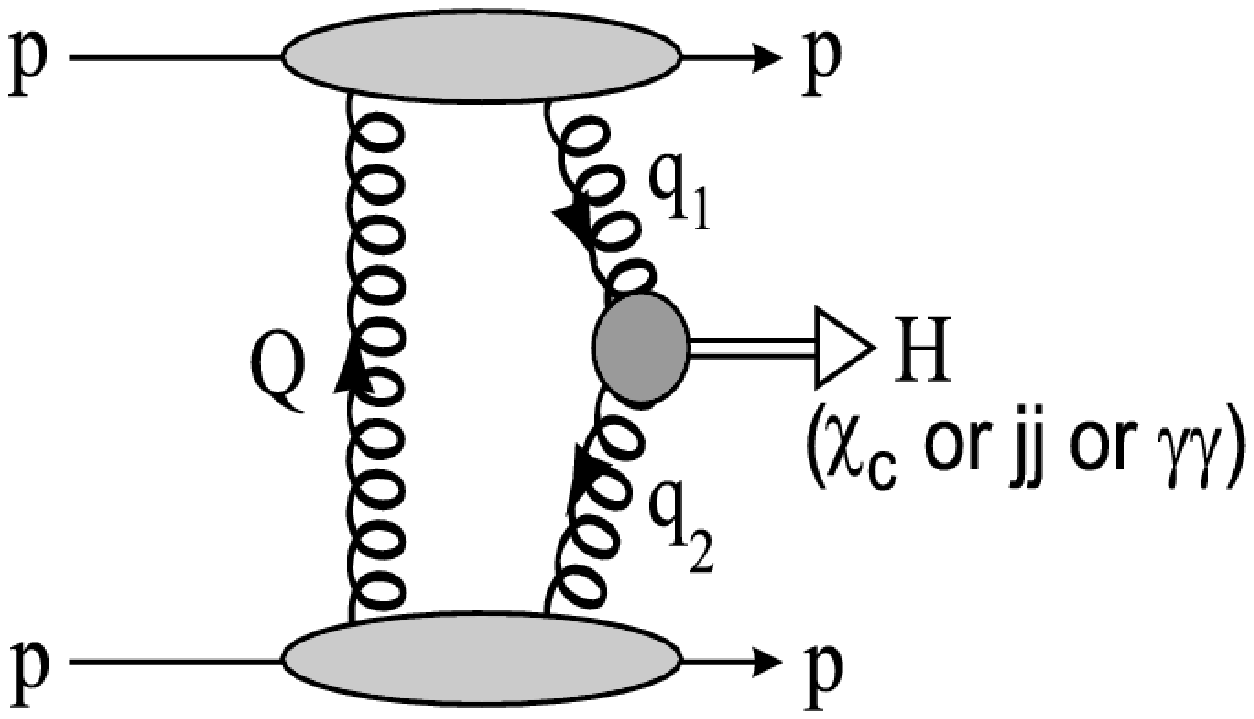}
\includegraphics[width=0.51\columnwidth,height=5cm]{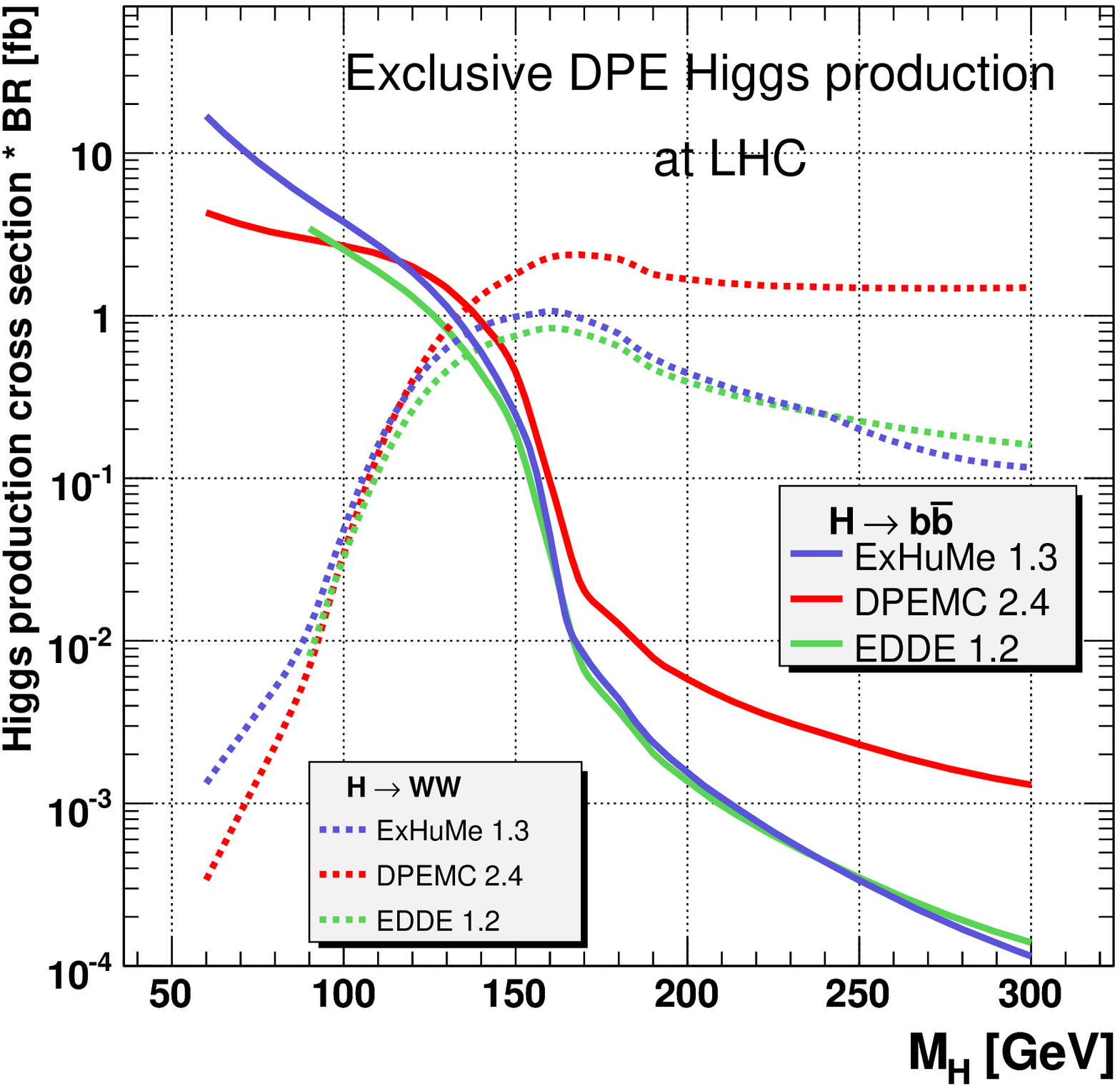}
}
\caption{Left: Central exclusive Higgs production 
via two-gluon exchange. Right: Cross sections for the SM Higgs (b$\barr{b}$, $WW$
channels) from various perturbative calculations~\cite{Albrow:2006xt}.}
\label{fig:DPE}
\end{figure}

\noindent
Central exclusive processes (CEP, Fig.~\ref{fig:DPE} left) are defined as 
pp $\rightarrow$ p $\oplus$ X $\oplus$ p where X is a fully measured simple 
state such as $\chi_{c,b}$, jet-jet ($j\,j$), $\lele$, $\gaga$, H, $W^+W^-$, ...
and '$\oplus$' represents a large rap-gap ($\Delta\eta\gtrsim$ 4). Central exclusive Higgs 
production, in particular, has attracted an important experimental and theoretical attention~\cite{fp420b,Khoze:2007td}. 
First, the expected SM cross sections are of order (3-10) fb (Fig.~\ref{fig:DPE}, right)
but, in minimal supersymmetric extensions of the SM (MSSM), can be a factor of 
10-100 larger  depending on $\tan{\beta}$.
Second, precise measurements of the proton momenta ($dp/p \approx 10^{-4}$) 
allow one to measure the Higgs mass with $\sigma(m_H)$~$\approx$~2~GeV, 
independent of the (central) decay mode (e.g. b$\barr{b}$, $WW$, $ZZ$). Third, 
spin selection rules suppress a large fraction of QCD production resulting in a very 
favourable 1:1 signal-to-background. 
Fourth, due to CEP J$^{\mbox{\tiny PC}}$ = 0$^{++}$ selection rules, 
azimuthal correlations of the outgoing protons are likely to provide the only method at 
hand at the LHC to easily determine the Higgs quantum numbers. Given the currently 
preferred range of Higgs masses, $m_H<$ 200 GeV, the optimal proton tagging 
acceptance is however beyond the current reach of TOTEM or ALFA. The FP420 R\&D 
collaboration proposes novel technologies (moving beam-pipe, 10-ps \v{C}erenkov 
detectors, ...) as ATLAS and CMS upgrades for proton tagging at $\pm$420 m.

\section{Low-$x$ QCD physics}

\begin{figure}[htbp]
\centerline{
\includegraphics[width=0.49\columnwidth,height=6.3cm]{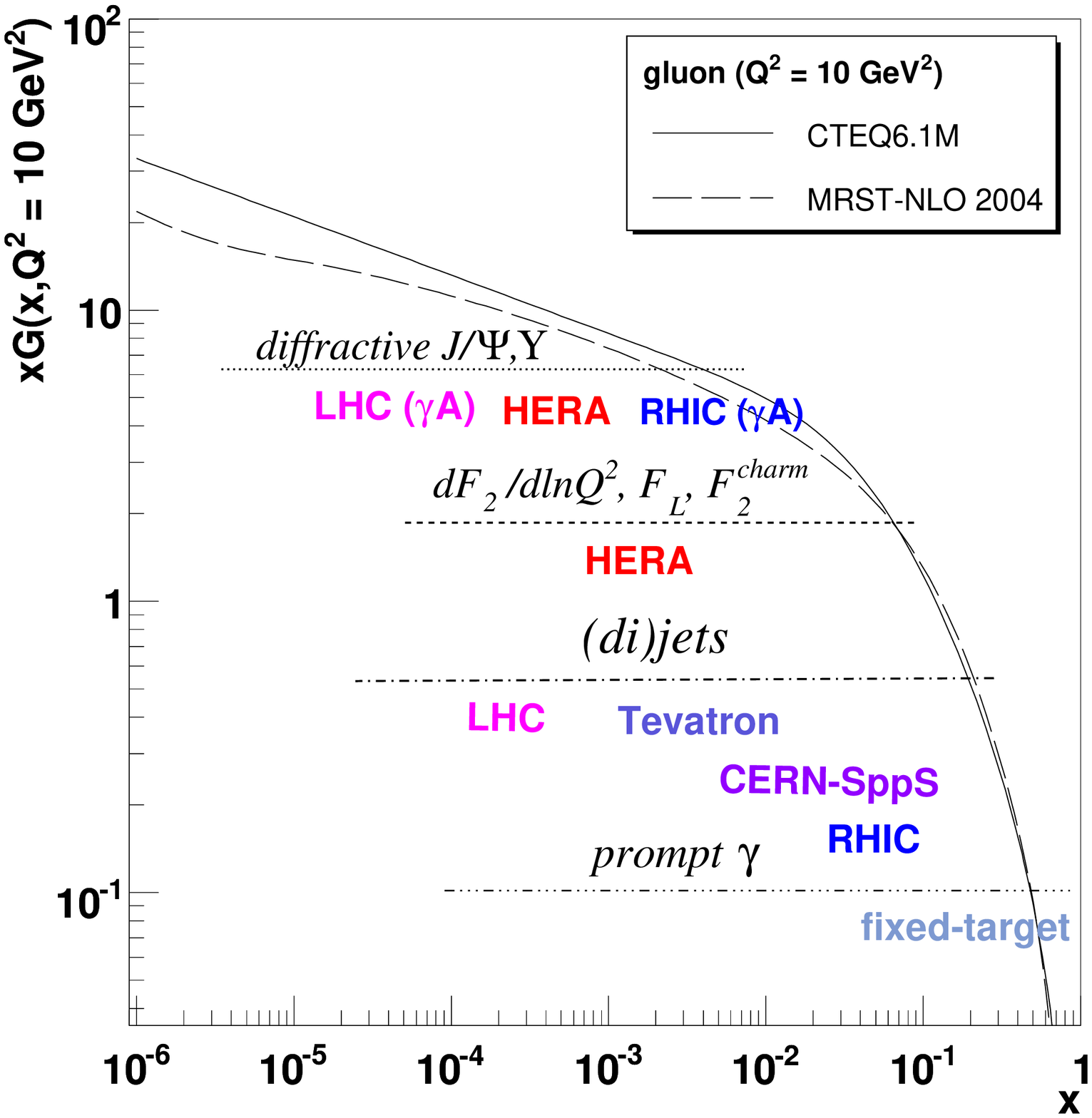}
\includegraphics[width=0.49\columnwidth,height=6.3cm]{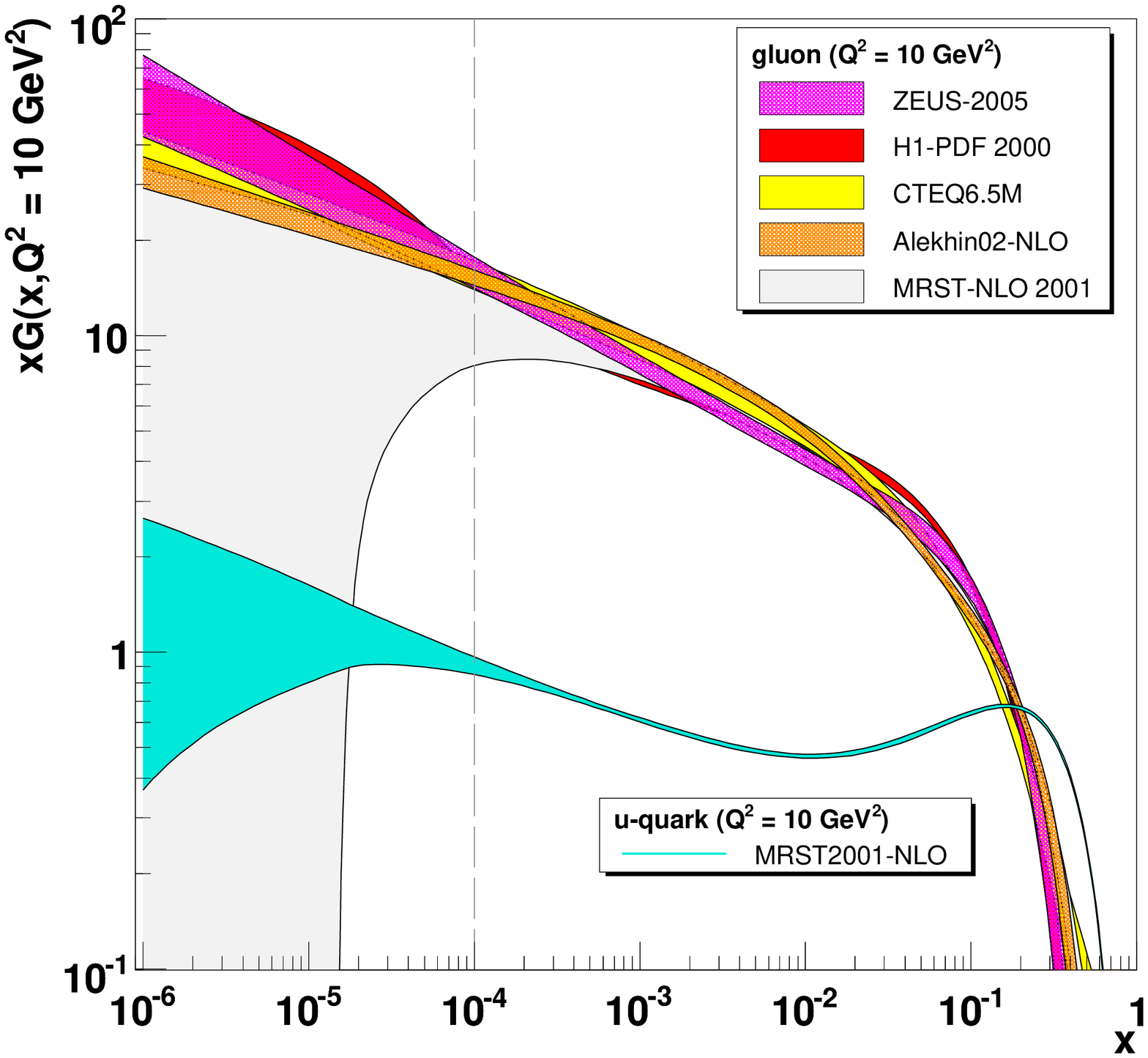}
}
\caption{Left: Experimental measurements of the gluon PDF. Right: Comparison of 
various fits~\protect\cite{pdf} of the proton $xG(x,Q^2$=10 GeV$^2)$ 
(the $u$ quark PDF is also shown, for reference).} 
\label{fig:HERA_xG}
\vspace{-.3cm}
\end{figure}
\begin{figure}[htbp]
\vspace{.3cm}
\centerline{
\includegraphics[width=6.5cm,height=4.8cm]{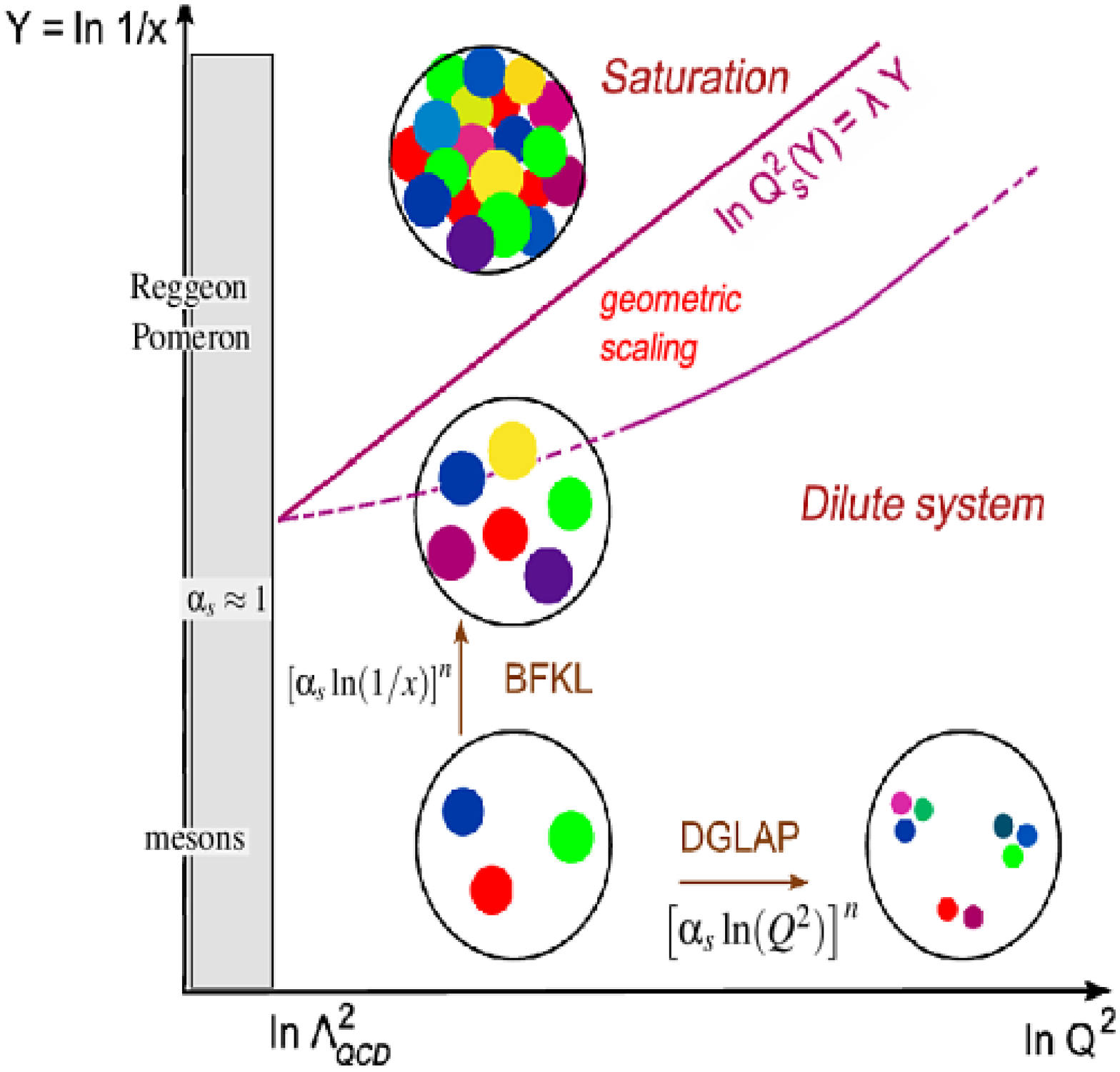}
\hspace{.3cm}
\includegraphics[width=7.3cm,height=5.cm]{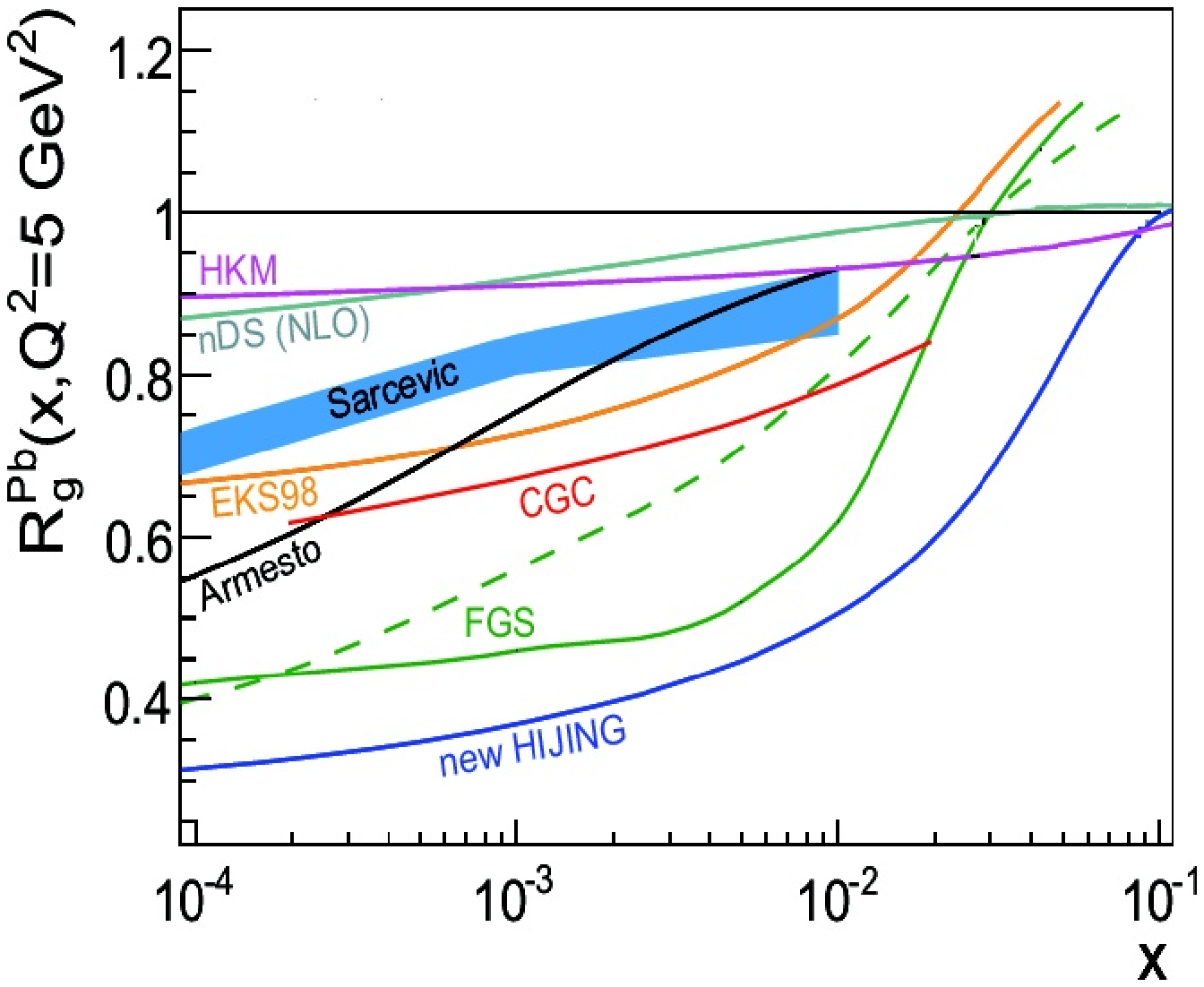}
}
\caption{Left: QCD log(1/$x$)-Q$^2$ plane with the different parton evolution regimes 
(DGLAP, BFKL, saturation). Right: Ratios of the Pb-over-proton gluon densities versus $x$ 
at fixed $Q^2$~=~5~GeV$^2$ from various nuclear shadowing parametrizations~\cite{armesto_shadow}.}
\label{fig:xG_Pb}
\end{figure}

\noindent
One of the main HERA observations is that the proton structure function is almost 
purely gluonic for values of the fractional momenta 
$x=p_{\mbox{\tiny{\it parton}}}/p_{\mbox{\tiny{\it proton}}} \lesssim 0.01$.
Fig.~\ref{fig:HERA_xG} summarises our current knowledge of the gluon density 
$xG(x,Q^2)$ in the proton. In DIS, the main source of information so far on $xG(x,Q^2)$ 
is (indirectly) obtained from the slope of the $F_2$ scaling violations. Additional 
constraints can be obtained from $F_2^{charm}$~\cite{hera_lhc_heavyQ}, 
and diffractive photoproduction of heavy vector mesons ($\jpsi,\ups$)~\cite{teubner07}.
The most direct access will come, however, from the longitudinal structure function $F_L$ 
whose measurement has driven the last (lower energy) runs at HERA~\cite{maxklein}.
In hadron-hadron collisions, $xG$ enters directly at LO in processes with prompt $\gamma$, 
jets, and  heavy-quarks in the final state. Below $x\approx$~10$^{-4}$ (10$^{-2}$) 
the gluon PDF in the proton (nucleus) is however poorly constrained as can be seen 
in the right plot of Fig.~\ref{fig:HERA_xG} (Fig.~\ref{fig:xG_Pb}). In this small-$x$
regime one expects non-linear gluon-gluon fusion processes -- not accounted 
for in the standard DGLAP/BFKL evolution equations -- to become important 
and tame the rise of the parton densities~\cite{raju} (Fig.~\ref{fig:xG_Pb} left). 
Such saturation effects are amplified in nuclear targets thanks to their increased 
transverse parton density. 

\noindent
Forward instrumentation provides an important lever arm for the measurement of the low-$x$ 
structure and evolution of the parton densities. Indeed, in a $2\rightarrow 2$ parton scattering 
the {\it minimum} momentum fraction probed when a particle of momentum $p_T$ is
produced at pseudo-rapidity $\eta$ is
\begin{equation}
x_{min} = \frac{x_T\,e^{-\eta}}{2-x_T\,e^{\eta}}\;\; \mbox{ where } \;\; x_T=2p_T/\sqrt{s}\,,
\label{eq:x2_min}
\end{equation}
i.e. $x_{min}$ decreases by a factor of $\sim$10 every 2 units of rapidity. 
Four representative measurements of the low-$x$ PDFs at the LHC are discussed 
next~\cite{d'Enterria:2006nb}.

\subsubsection*{$\bullet$ Case study I: Forward (di)jets}
\begin{figure}[htbp]
\centerline{
\includegraphics[width=0.49\columnwidth,height=5.8cm]{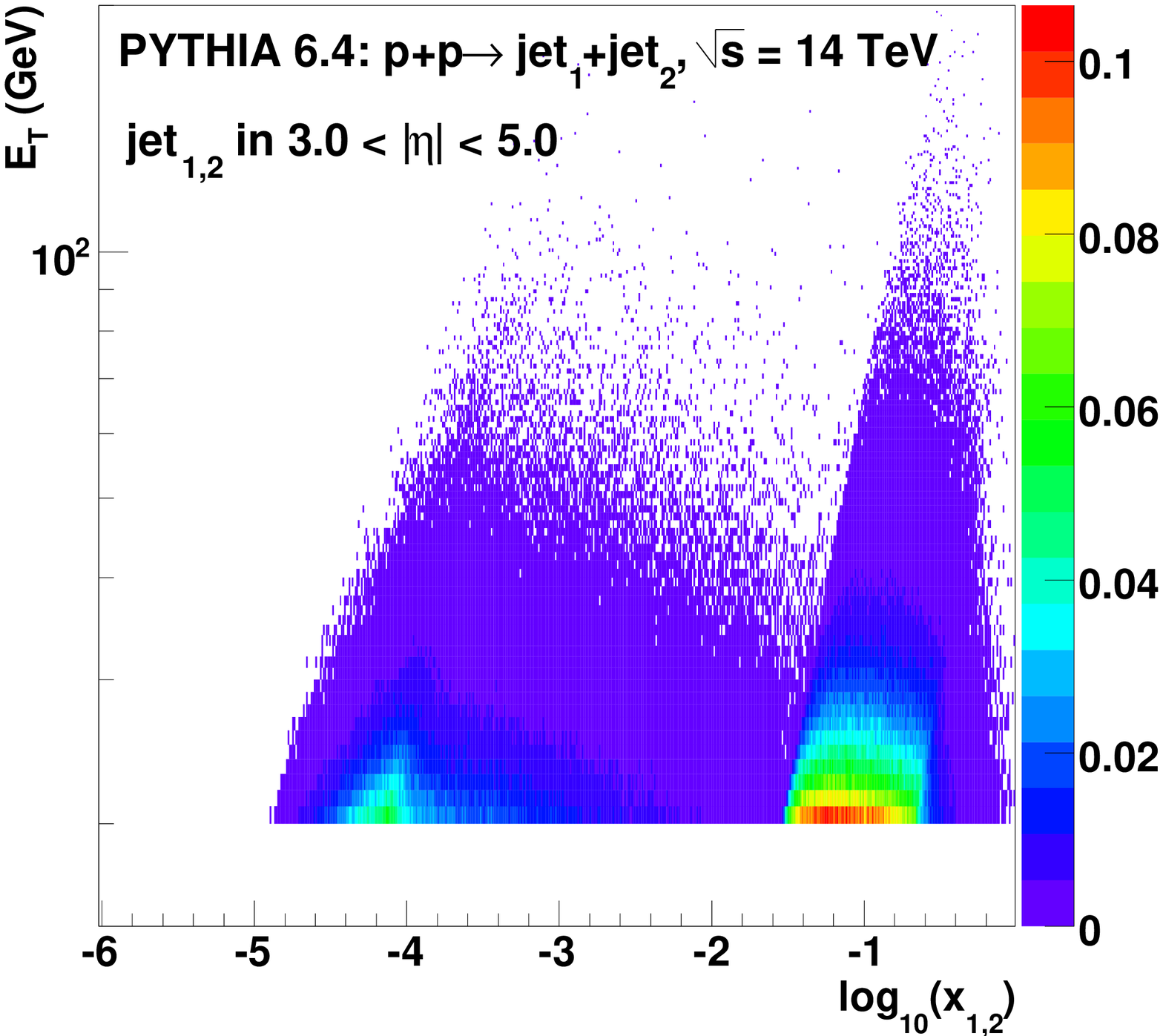}
\hspace{1cm}
\includegraphics[width=0.49\columnwidth,height=5.5cm]{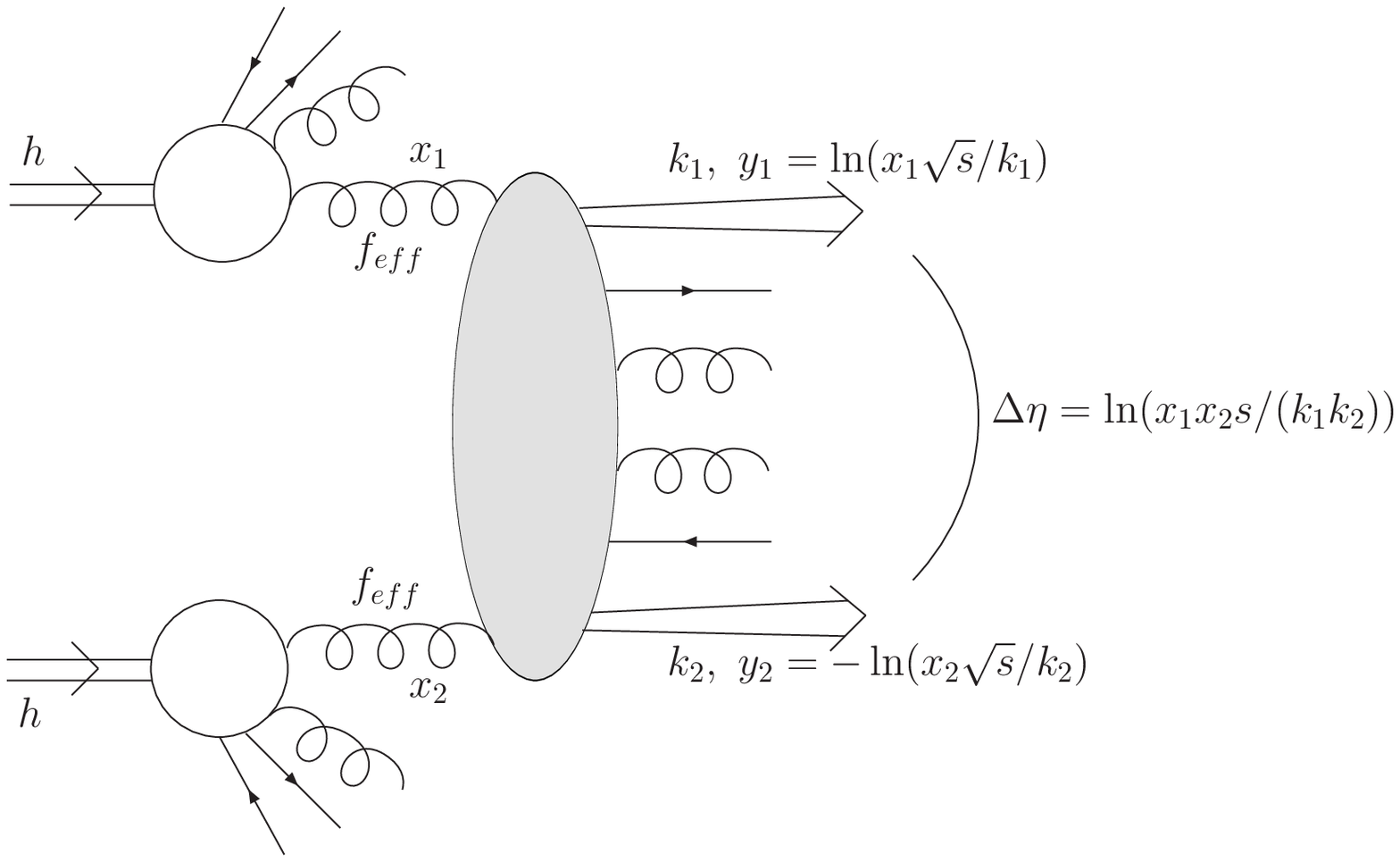}
}
\caption{Left: Parton $x_{1,2}$ distributions probed in p-p collisions at $\sqrts$ = 14 TeV 
with single jet production within ATLAS/CMS forward calorimeters acceptances.
Right: M\"uller-Navelet dijet production diagram in p-p collisions.}
\label{fig:fwd_jets}
\end{figure}

\noindent
The measurement of (relatively soft) jets with  $p_T\approx$ 20 - 100 GeV/$c$ in p-p 
at 14 TeV in the CASTOR forward calorimeter (5.2$<|\eta|<$6.6) allows one to probe 
the PDFs at $x$ values as low as $x\approx 10^{-6}$ (see Fig.~\ref{fig:fwd_jets} left, 
for jets in ATLAS FCal and CMS HF calorimeters). 
In addition to the single inclusive cross sections, the production of events with two
similar transverse-momentum jets emitted in each one of the forward/backward directions, 
the so-called ``M\"uller-Navelet jets'' (Fig.~\ref{fig:fwd_jets} right), is a particularly 
sensitive measure of BFKL~\cite{mueller_navelet} as well as non-linear~\cite{marquet05} 
parton evolutions. The large rapidity interval between the jets (e.g. $\Delta\eta \approx$ 10 
in the extremes of HF+ and HF-) enhances large logarithms of the type 
$\Delta\eta \sim log(s/k_1k_2)$ which can be appropriately resummed using the BFKL equation.
The phenomenological consequences expected in the BFKL regime are enhanced M\"uller-Navelet
dijet rates and wider azimuthal decorrelations for increasing $\Delta\eta$ 
separations~\cite{Royon:2007aj,Vera:2007ba}. Preliminary CMS analyses~\cite{Albrow:2006xt} 
indicate that such studies are well feasible measuring jets in each one of the hadronic forward 
(HF) calorimeters.

\subsubsection*{$\bullet$ Case study II: Forward heavy-quarks}
\label{sec:heavyQ}
\begin{figure}[htbp]
\centerline{
\hspace*{-1.3cm}
\includegraphics[width=0.49\columnwidth,height=5.9cm]{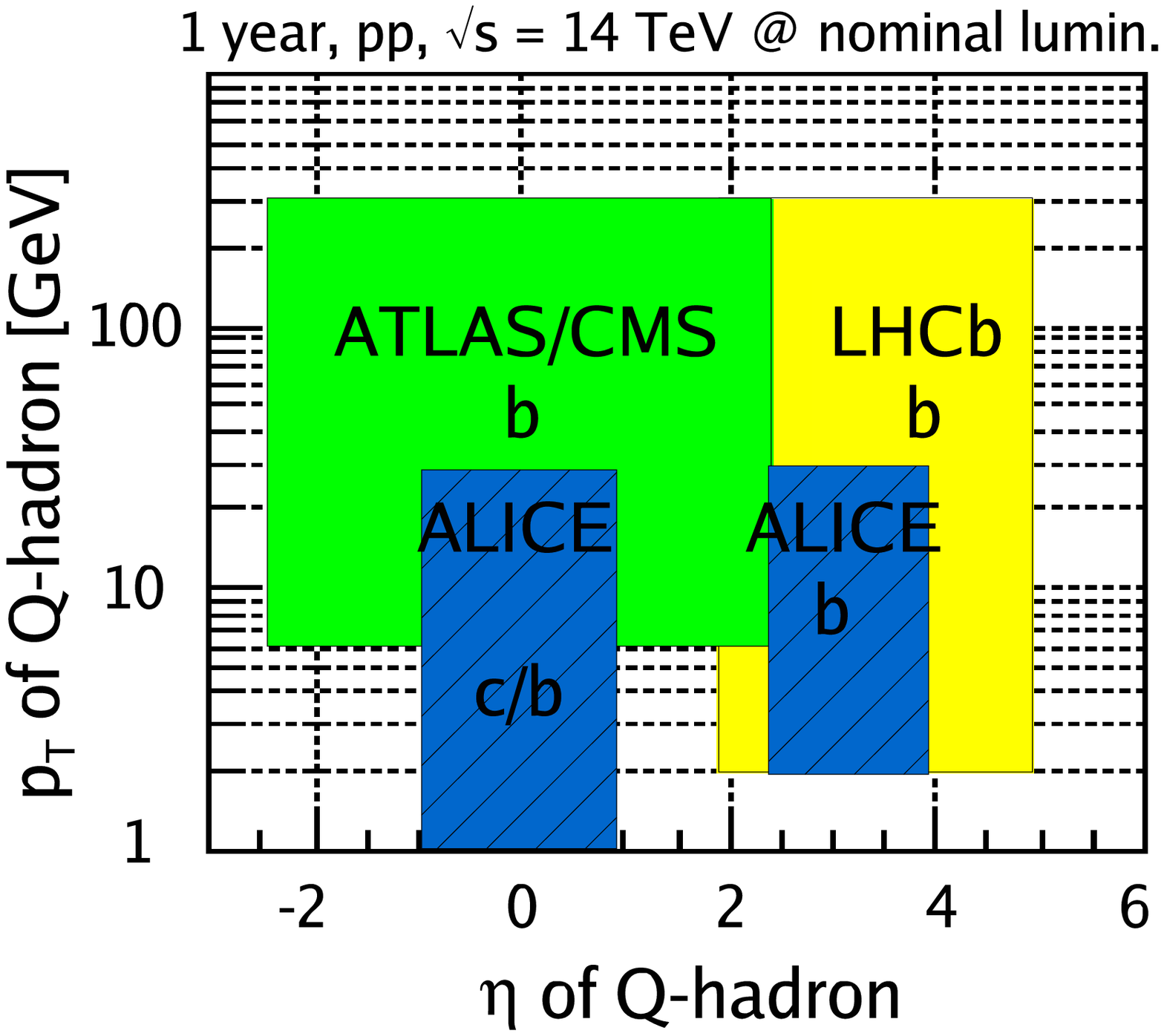}
\hspace*{.1cm}
\includegraphics[width=5.5cm,height=5.5cm,angle=90]{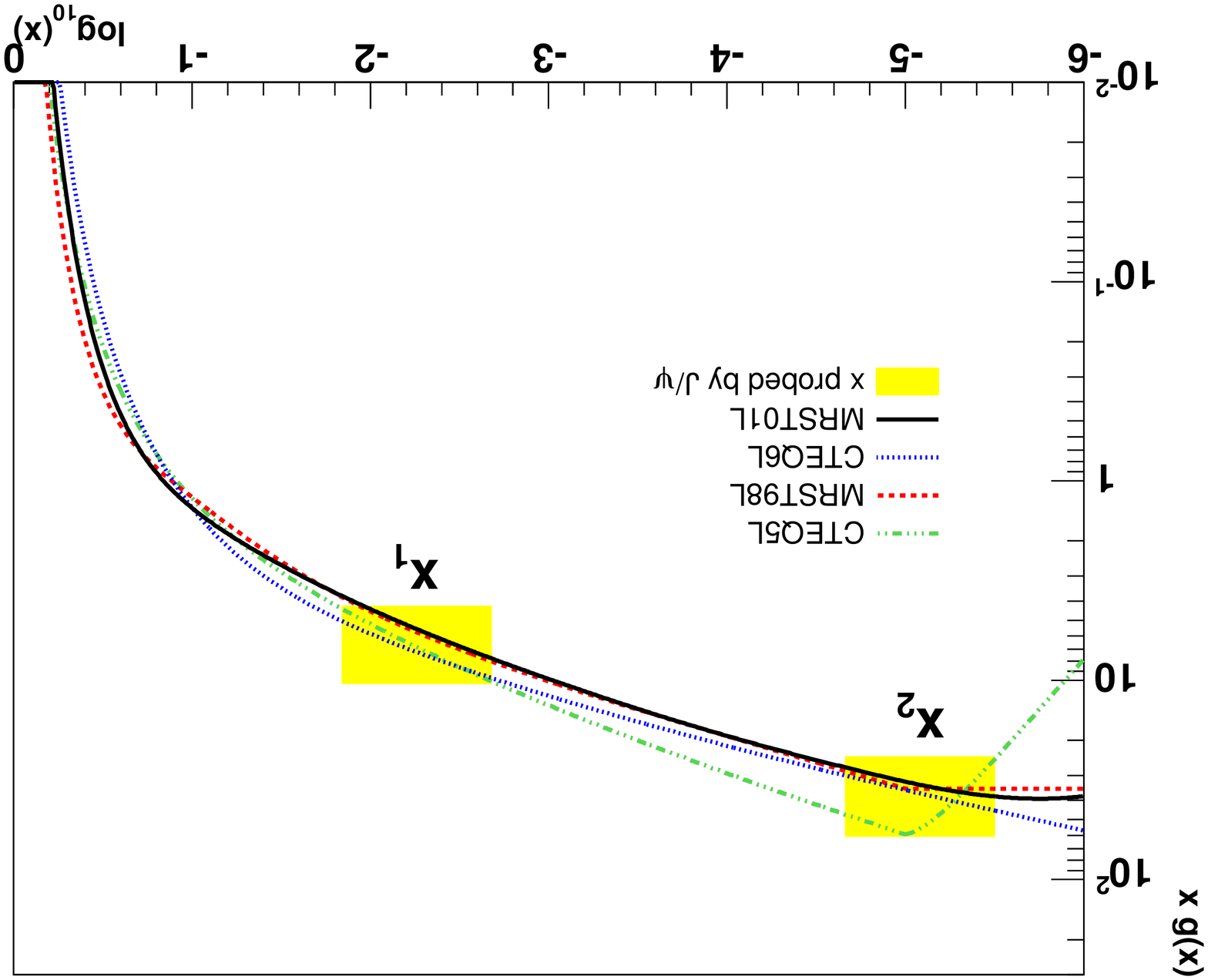}
}
\caption{Left: Acceptances in $(\eta,p_T)$ for open charm and bottom at the LHC~\cite{hera_lhc_heavyQ}. 
Right: Sensitivity of the forward $\jpsi$ measurement in ALICE to the gluon PDF~\cite{stocco}.}
\label{fig:heavyQ_QQbar_LHC}
\end{figure}

The possibility of ALICE and LHCb (Fig.~\ref{fig:heavyQ_QQbar_LHC}, left) to reconstruct 
heavy D and B mesons as well as quarkonia in a large forward rapidity range can also
put stringent constraints on the gluon structure and evolution at low-$x$.
Studies of small-$x$ effects on heavy flavour production based on collinear and 
$k_T$ factorization, including non-linear terms in the parton evolution, lead to varying 
predictions for the measured $c$ and $b$ cross sections at the LHC~\cite{hera_lhc_heavyQ}. 
The hadroproduction of $\jpsi$ proceeds mainly via gluon-gluon fusion and, having a mass 
around the saturation scale $Q_s^{\mbox{\sc lhc}}\approx$ 3 GeV, 
is also a sensitive probe of possible gluon saturation phenomena. 
Figure~\ref{fig:heavyQ_QQbar_LHC} right, shows the gluon $x$ range probed in p-p collisions 
producing a $\jpsi$ inside the ALICE muon arm acceptance ($2.5\lesssim \eta \lesssim 4$). 
The observed differences in the underlying PDF fits translate into variations as large as 
a factors of $\sim$2 in the finally measured cross sections~\cite{stocco}.

\subsubsection*{$\bullet$ Case study III: Q$\barr{Q}$ exclusive photoproduction}
\label{sec:upc_qqbar}
\begin{figure}[htbp]
\centerline{
\includegraphics[width=0.49\columnwidth,height=4.5cm]{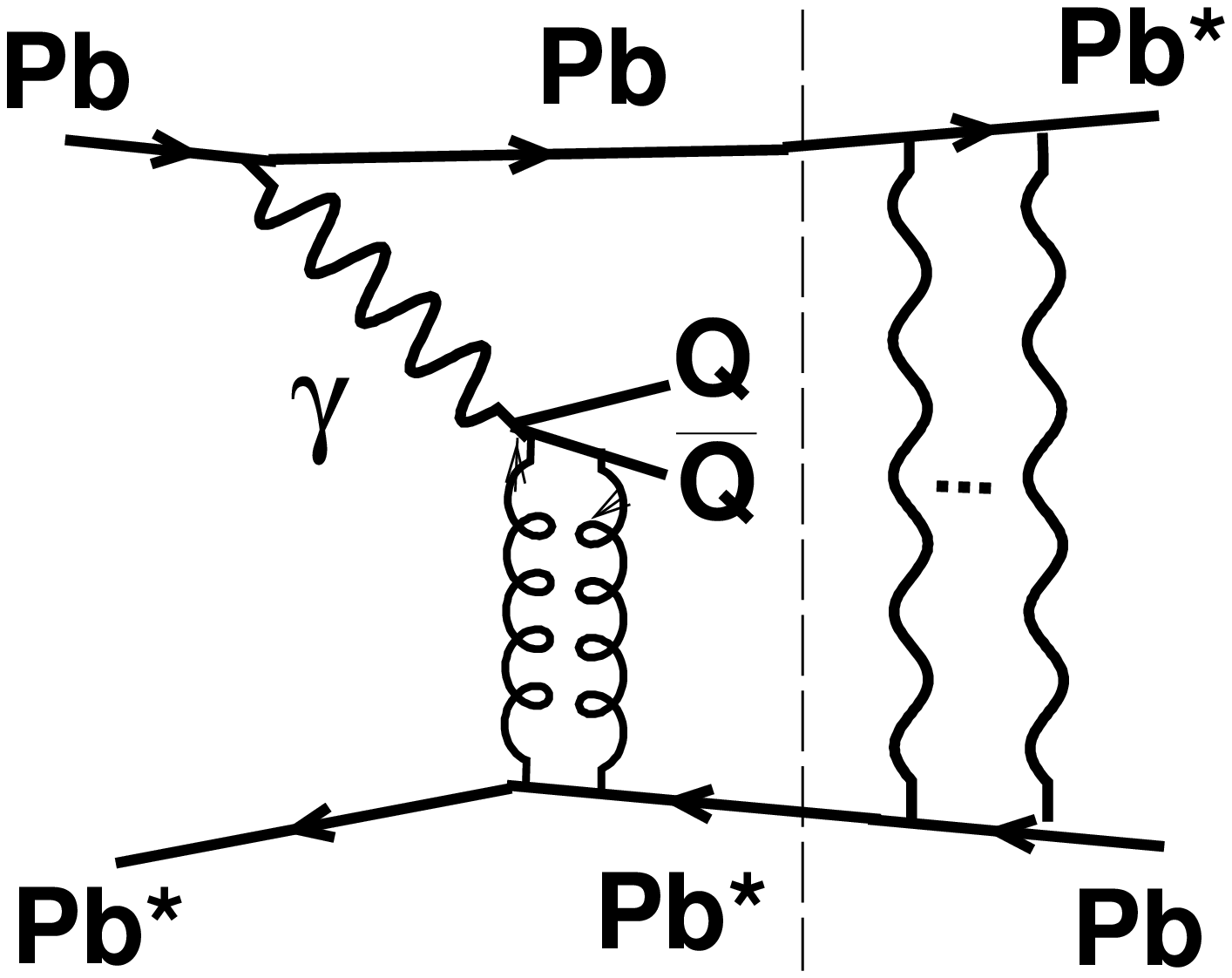}
\includegraphics[width=0.49\columnwidth,height=5cm]{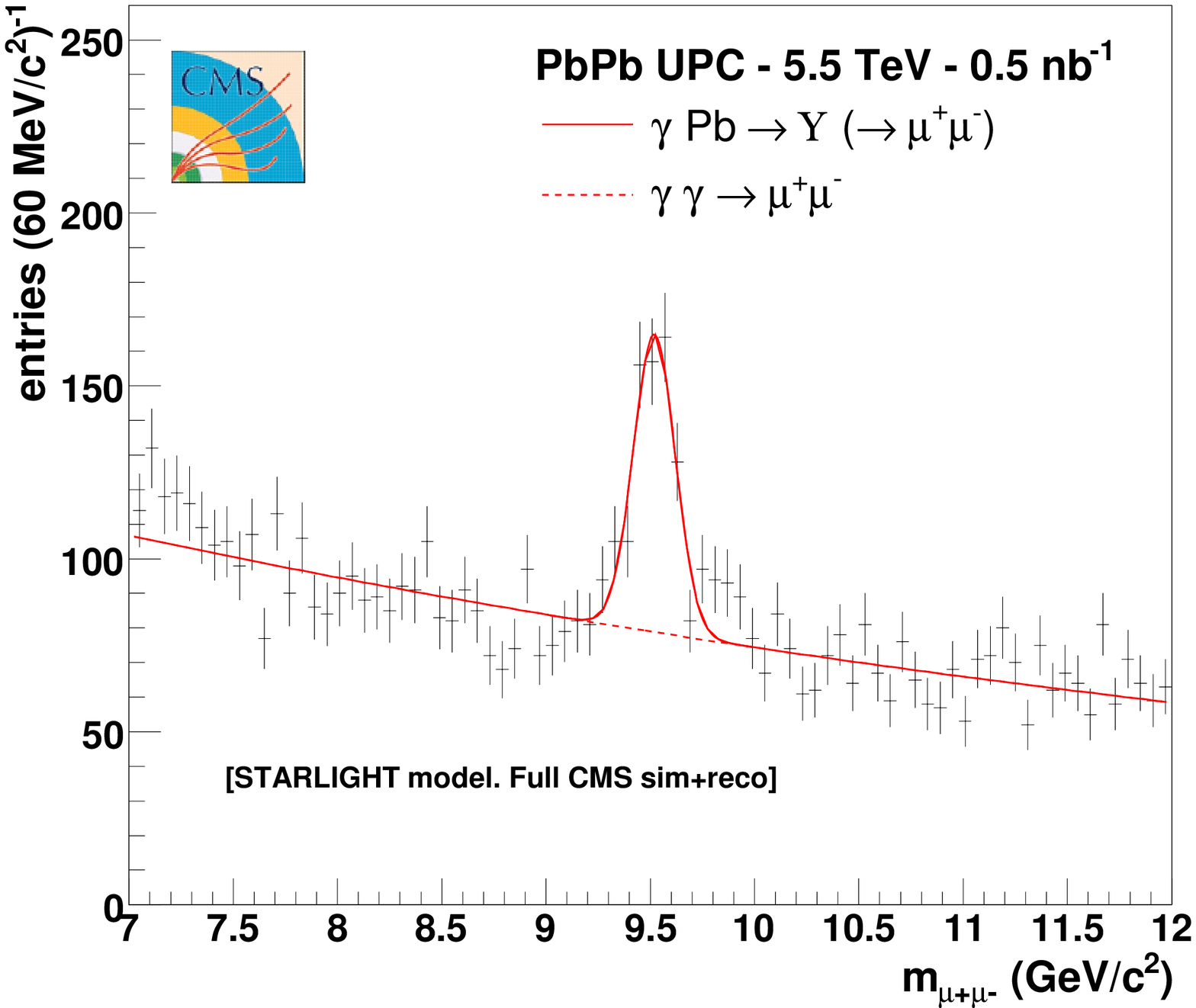}
}
\caption{Left: Exclusive quarkonia photoproduction in UPCs.
Right: Expected dimuon invariant mass from $\gamma\,$Pb$\rightarrow \Upsilon\,$Pb$^\star$ 
on top of $\gaga\rightarrow \mumu$ continuum in UPC Pb-Pb at 5.5 TeV~\protect\cite{cms_hi_ptdr}.}
\label{fig:upc_ups_cms}
\end{figure}

\noindent
Ultra-peripheral interactions (UPCs) of high-energy heavy ions generate strong electromagnetic fields 
which can be used to constrain the low-$x$ behaviour of the nuclear gluon density via exclusive 
photoproduction of quarkonia, dijets and other hard processes~\cite{Baltz:2007kq}.
Lead beams at 2.75 TeV have Lorentz factors $\gamma$ = 2930 leading to maximum (equivalent) 
photon energies $\omega_{\ensuremath{\it max}}\approx \gamma/R\sim$ 100 GeV,
and corresponding maximum c.m. energies: 
$W_{\gaga}^{\ensuremath{\it max}}\approx$ 160 GeV and $W^{\ensuremath{\it max}}_{\gA}\approx$ 1 TeV,
i.e. 3--4 times higher than equivalent photoproduction studies at HERA.
The $x$ values probed in $\gamma\,$A$\rightarrow$Q$\barr{Q}\;$A processes (Fig.~\ref{fig:upc_ups_cms}, left) 
can be as low as $x\sim 10^{-5}$~\cite{Baltz:2007kq}. The ALICE, ATLAS and CMS experiments 
can measure the $\jpsi,\ups\rightarrow e^+e^-,\mu^+\mu^-$ produced in electromagnetic 
Pb-Pb collisions tagged with neutrons detected in the ZDCs, as done at RHIC~\cite{dde_qm05}. 
Full simulation+reconstruction analyses~\cite{cms_hi_ptdr} indicate that CMS can measure 
a total yield of $\sim$\,500 $\ups$'s within $|\eta|<$ 2.5 for 0.5 nb$^{-1}$ nominal Pb-Pb integrated 
luminosity (Fig.~\ref{fig:upc_ups_cms}). With such statistics, studies of the $p_T$ and $\eta$ 
distributions of the $\ups$ can be carried out which will help to constrain the low-$x$ gluon 
density in the Pb nucleus.

\subsubsection*{$\bullet$ Case study IV: Forward Drell-Yan pairs} 
\noindent
High-mass Drell-Yan pair production at the very forward rapidities covered by LHCb and by the 
CASTOR and TOTEM T2  detectors can probe the parton densities down to $x=M/\sqrts \;e^{\pm y}\sim10^{-6}$
at higher virtualities $M^2$ than those accessible in other measurements discussed here.
A study is currently underway in CMS~\cite{Albrow:2006xt} to combine the CASTOR 
electromagnetic energy measurement together with the good position resolution of T2 for charged 
tracks, to trigger on and reconstruct $\elel$ pairs in p-p collisions at 14 TeV,
and scrutinise $xg$ in the $M^2$ and $x$ plane.

\section{Cosmic-rays physics connection}

\begin{figure}[htbp]
\centerline{
\includegraphics[width=0.49\columnwidth,height=5.5cm]{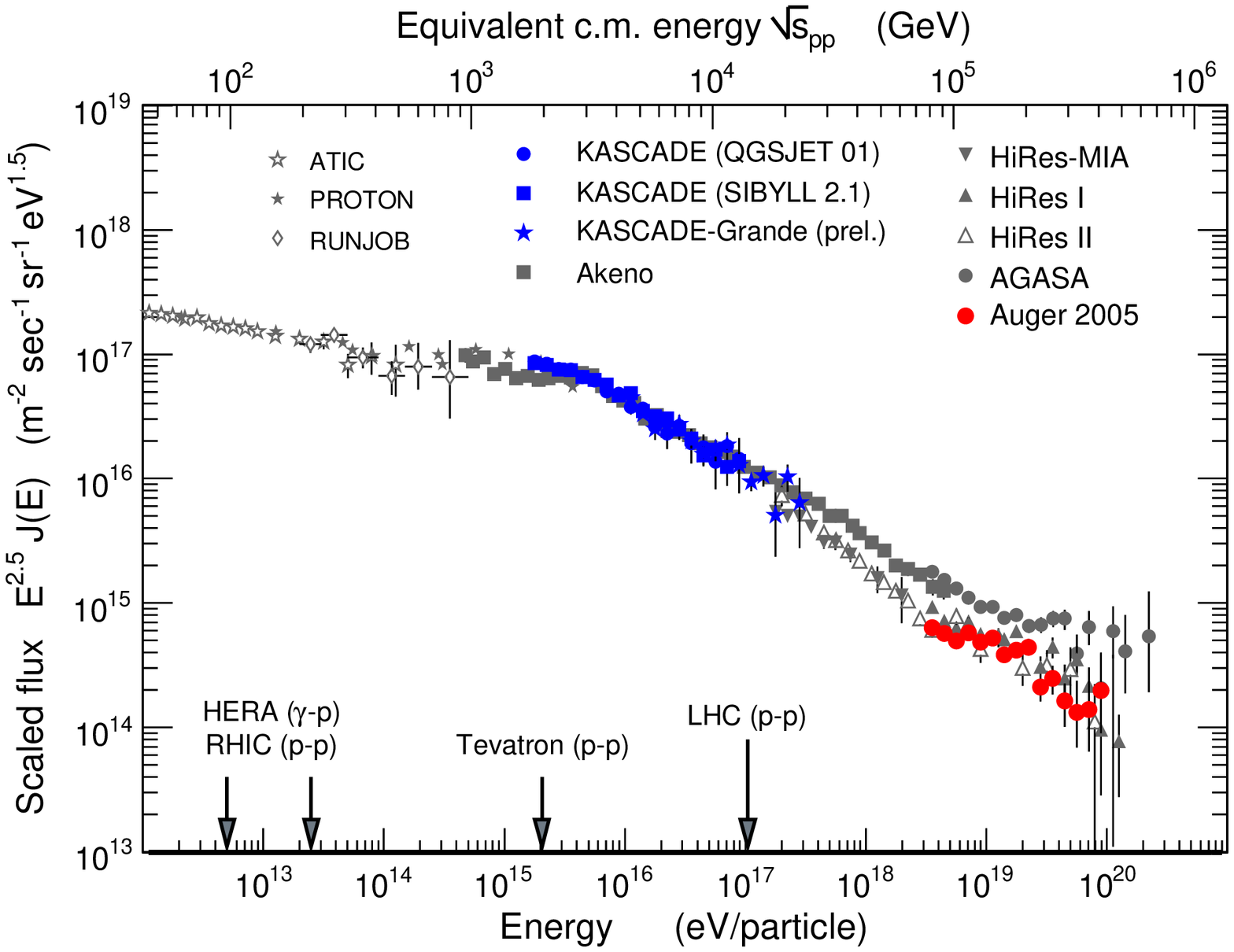}
\includegraphics[width=0.49\columnwidth,height=5.1cm]{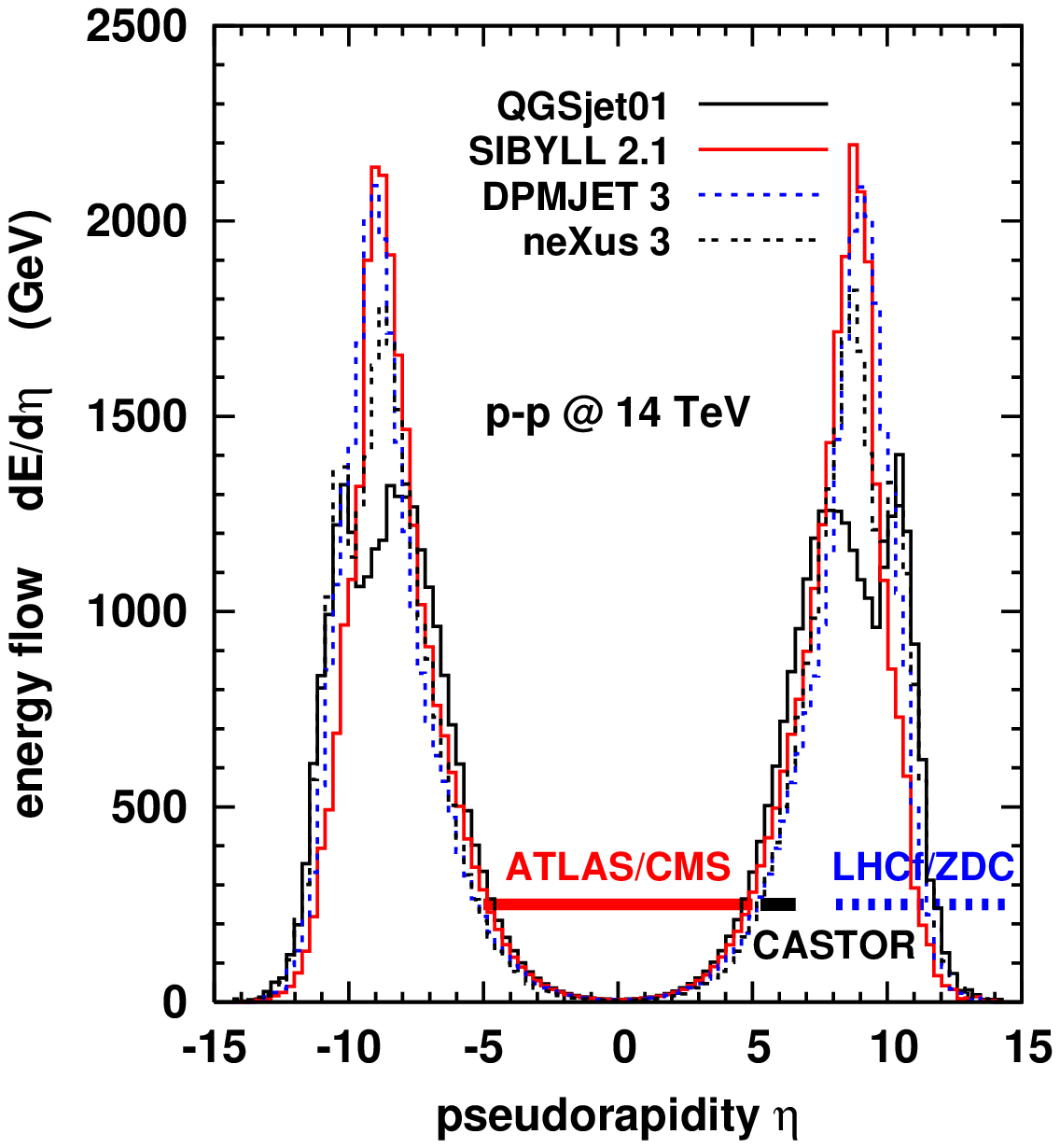}
}
\caption{Left: The cosmic-ray energy distribution~\protect\cite{Engel:2005gf}. 
Right: Pseudo-rapidity energy distribution for p-p at the LHC predicted by four commonly 
used MC models in UHE cosmic rays physics (the acceptance of LHCf and ZDCs refers to
neutral particles)~\protect\cite{Albrow:2006xt}.}
\label{fig:cosmics}
\end{figure}

\noindent
The origin of cosmic rays (CRs) with energies above $10^{15}$\,eV is unclear, as it is 
the identity of the primaries. Due to their low fluxes (less than 1 particle per m$^2$ and year, 
see Fig.~\ref{fig:cosmics} left) 
only indirect measurements exist which use the atmosphere as a ``calorimeter''. 
The energy and mass of UHE cosmic rays are then obtained with the help of Monte Carlo 
(MC) codes which describe the shower development (dominated by forward and soft QCD 
interactions) in the upper atmosphere~\cite{Engel:2005gf}. The existing 
MC models (Fig.~\ref{fig:cosmics}, right) predict energy and multiplicity flows 
differing by factors as large as three, with significant inconsistencies in the 
forward region ($|\eta| >$ 5). Forward measurements at LHC energies 
($E_{lab} \approx 10^{17}$ eV) in p-p, p-A and A-A collisions\footnote{Note that CRs interactions 
in the atmosphere are mostly proton-nucleus (p-Air) and nucleus-nucleus ($\alpha$-,Fe- Air) collisions.} 
will provide strong constraints to calibrate and tune these models and make more reliable 
predictions for the CR energy and composition at the highest energies observed.
Forward measurements at the LHC, especially in calorimeters with longitudinal segmentation 
like CASTOR, will in addition help to interpret exotic CR topologies like the so-called
``Centauro'' events~\cite{castor}. 

\section{Electroweak physics}

\begin{figure}[htbp]
\centerline{
\includegraphics[height=3.cm]{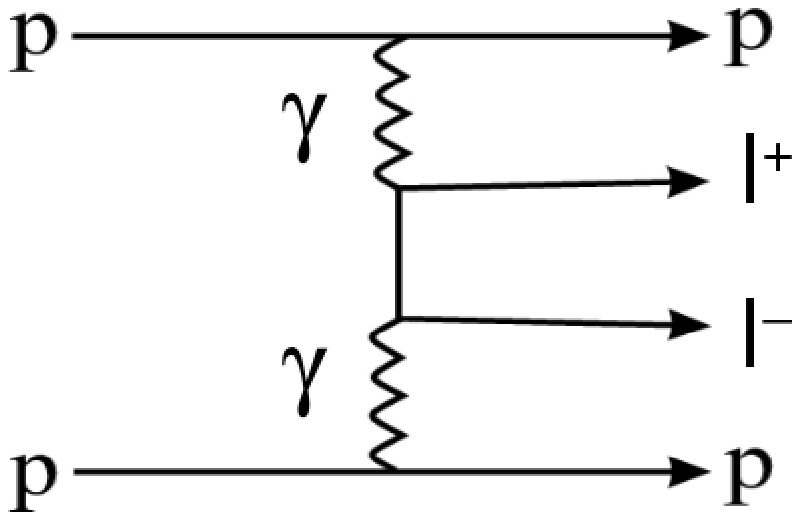}
\includegraphics[height=3.cm]{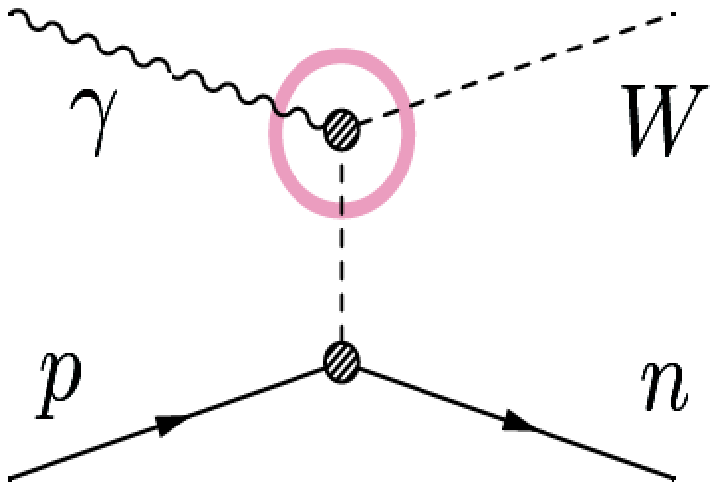}
}
\caption{Photoproduction diagrams in electromagnetic proton-proton
interactions: two-photon dilepton production (left), and single-W photoproduction (right).
In both processes, the forward-going protons (neutrons) can be detected in RPs (ZDCs).}
\label{fig:gg_gp}
\end{figure}

\noindent
Interesting electroweak processes in photon-photon and photon-proton,-nucleus interactions, 
tagged with forward instrumentation, will be also accessible for the first time at TeV energies at 
the LHC. Two-photon dilepton production, pp $\rightarrow$ p$\;\lele$p (Fig.~\ref{fig:gg_gp}, left) 
will be an excellent luminosity calibration process, with a very well known QED 
cross section~\cite{Bocian:2004ev}. Experimentally, such a process can be tagged with forward
protons and has a clear signature in the exclusive back-to-back dielectrons (dimuons) measured 
e.g. in CASTOR/T2 (in the central muon chambers). The p-p cross section calculated using 
{\sc LPAIR} for events where both muons have $p_T >$ 3 GeV/$c$ and can, therefore, 
reach the CMS muon chambers is about 50 pb. About 300 events per 100 pb$^{-1}$ are 
thus expected in CMS after muon trigger cuts~\cite{Albrow:2006xt}. The situation is much 
more favourable in the case of Pb-Pb collisions since the dilepton continuum is much larger 
than in p-p ($Z^4$ enhancement factor, see Fig.~\ref{fig:upc_ups_cms} right) and the forward 
neutron tagging much more efficient than the forward proton one.\\ 

\noindent
The couplings of gauge bosons among themselves belong to one of the least tested sectors 
of the electroweak theory. A process well-suited to testing the ($ WW \gamma$) gauge 
boson self-interaction is the photoproduction of single $W$ bosons from a nucleon 
(Fig.~\ref{fig:gg_gp}, right) in ultra-peripheral p-p~\cite{Ovyn07}, p-A and A-A~\cite{Baltz:2007hw}
collisions. A large cross section of about 1 pb is expected for large photon-proton c.m. 
energies, $W_{\gamma\,p}>$ 1 TeV. In addition, the two-photon $W^+W^-$ 
exclusive production probes {\it quartic} gauge-boson-couplings. The process has a total 
cross section of more than 100 fb, and a very clear signature. Its cross section is still about 
10 fb for $W_{\gamma\,p}>$ 1 TeV showing sensitivity to physics beyond the SM~\cite{Ovyn07}. 

\section*{Acknowledgments}

\noindent
I would like to thank the organisers of DIS'07 -- and, in particular, the
conveners of the {\em Future of DIS} Section -- for their kind invitation to 
such a stimulating conference. Special thanks due to Albert de Roeck,
Michele Arneodo and Monika Grothe for valuable discussions. Work 
supported by the 6th EU Framework Programme contract MEIF-CT-2005-025073.

%

\def\IJMPA{{Int. J. Mod. Phys.}~{\bf A}}
\def\EPJ{{Eur. Phys. J.}~{\bf C}}
\def\JPG{{J. Phys.}~{\bf G}}
\def\JHEP{{J. High Energy Phys.}~}
\def\NCA{Nuovo Cimento~}
\def\NIM{Nucl. Instrum. Methods~}
\def\NIMA{{Nucl. Instrum. Methods}~{\bf A}}
\def\NPA{{Nucl. Phys.}~{\bf A}}
\def\NPB{{Nucl. Phys.}~{\bf B}}
\def\PLB{{Phys. Lett.}~{\bf B}}
\def\PLC{Phys. Repts.\ }
\def\PRL{Phys. Rev. Lett.\ }
\def\PRD{{Phys. Rev.}~{\bf D}}
\def\PRC{{Phys. Rev.}~{\bf C}}
\def\ZPC{{Z. Phys.}~{\bf C}}

\begin{footnotesize}

\end{footnotesize}


\end{document}